\documentclass[showpacs,aps,prl,twocolumn]{revtex4}
\usepackage{graphicx}
\usepackage{amsfonts}

\begin{document}
\title{Propagation of matter wave solitons in periodic and random nonlinear potentials }
\author{Fatkhulla Kh. Abdullaev\dag \, and Josselin Garnier\ddag\footnote[7]{Corresponding author (garnier@math.jussieu.fr)}
}
\affiliation{\dag\
Dipartimento di Fisica "E.R. Caianiello", Universit\'a di Salerno, 84081 Baronissi (SA), Italy\\
\ddag\ Laboratoire de Probabilit\'es et Mod\`eles Al\'eatoires
\& Laboratoire Jacques-Louis Lions,
Universit{\'e} Paris VII,
2 Place Jussieu, 75251 Paris Cedex 5,
France}

\begin{abstract}
We study the motion of bright matter wave solitons in nonlinear
potentials, produced by periodic or random spatial variations of
the atomic scattering length. We obtain analytical results for the
soliton motion, the radiation of matter wave, and the radiative
soliton decay in such configurations of the Bose-Einstein
condensate. The stable regimes of propagation are analyzed. The
results  are in remarkable agreement with the numerical
simulations of the Gross-Pitaevskii equation with periodic or
random spatial variations of the mean field interactions.
\end{abstract}
\maketitle

\newpage

{\it Introduction.} Nonlinear excitations in Bose-Einstein
condensates (BEC) have attracted a lot of attention recently. In
particular matter wave solitons are interesting  from the
fundamental point of view \cite{Brazhnii}. The discovery of matter
waves solitons in BEC \cite{Burger,Straeker,Khaykovich,Eiermann}
has opened the possibility to develope new methods for generating
and controlling solitons. The investigation of the soliton
dynamics in inhomogeneous BEC is of interest, in particular with
inhomogeneities periodic in time or space. Time variations can be
achieved with the Feshbach resonance (FR) management technique and it
has been studied in \cite{Abd1,Abd2,Kevrek2,Theo}. New type of
solitons can be generated by this way, and stabilization of
higher-dimensional solitons in attractive condensate has been
shown. Spatial variations have been investigated in the form of a
periodic or random linear potential. In particular, a periodic
optical lattice can be produced by counter propagating laser
beams. Such potentials can be used to control the soliton
parameters \cite{Scharf,Kartashov} or to generate gap bright
solitons \cite{Eiermann}. Propagation in a random linear potential
was considered from the point of view of the Anderson localization
in BEC and the observation of the crossover between the Anderson
localization regime and the nonlinear regime \cite{Castin}. Dark
soliton propagation in a linear random potential was studied in
\cite{Pavloff}. In the recent work \cite{Malomed} the properties
of stationary localized states in the nonlinear optical lattice
has been investigated. The problems of matter wave soliton
propagation, when the mean field nonlinearity varies periodically
or randomly in space remain open.

In this Rapid Communication  we consider the propagation of
nonlinear matter wavepackets and waves emission in the presence of
a new type of inhomogeneities, namely under {\it nonlinear
periodic or random potential}, produced by periodic or random
variations of the atomic scattering length in space. The strength
and the sign of the interatomic interactions, i.e. the value and
sign of the atomic scattering length $a_s$, can be varied using
the FR method. Small variations of an external
magnetic field near the FR can lead to large
variations of $a_s$. For example if we consider the
one-dimensional Bose gas close to the magnetic wire, then by small
variations of the current one can induce spatially random magnetic
field fluctuations. This in turn generates random spatial
fluctuations of the strength of the interatomic interactions
\cite{Gimperlein,Wildermuth}. Such variations can be achieved also
by the optically induced FR \cite{FR1,FR2}.
The Gross-Pitaevskii (GP) equation describing such configurations
has a periodically or randomly varying in space mean field
nonlinear coefficient.

The dynamics of quasi-1D nonlinear matter waves is described by
the GP equation \cite{Abd2,Theo}
\begin{equation}\label{gp1}
i\hbar\psi_{t} = -\frac{\hbar^2}{2m}\psi_{xx}+
g_{1D}|\psi|^{2}\psi .
\end{equation}
Here $\psi$ is the mean field wavefunction, with $\int |\psi|^2 dx
= N$, $N$ is the number of atoms, $g_{1D} = 2\hbar
\omega_{\perp}a_{s}$, where $\omega_{\perp}$ is the transverse
oscillator frequency, $a_{s}(x)= a_{s0} + a_{s1}f(x)$ is the
spatially dependent  atomic scattering length. The spatial
dependence will be assumed to be periodic or random. In
dimensionless variables where the distance $x$ is measured in
units of the healing length $\xi = \hbar/\sqrt{n_{0}g_{1D}m}$,
with $n_{0}$ the peak density,  and the time $t$ is measured in
$t_{0} = \xi/(2c)$, where $c = \sqrt{n_{0}g_{1D}/m}$, we obtain
the equation
\begin{equation}\label{gp2}
iu_{t} + u_{xx} + 2|u|^{2}u = - V(x)|u|^{2}u.
\end{equation}

{\it Matter wave soliton motion in a nonlinear periodic
potential}. We consider the case when the
incident wave is the soliton incoming from the left:
\begin{equation}
u ^{(s)} (x,t) = 2\nu_{0}\frac{\exp i(2\mu_{0}(x-4\mu_{0}t) + 4(\nu_0^2
+ \mu_{0}^2 )t )}{\cosh(2\nu_{0}(x-4\mu_{0}t))},
\end{equation}
where $2\nu_{0}$ and $4\mu_{0}$ are the soliton amplitude and velocity,
respectively. The scattering length has spatial periodic
modulations, so we have $V(x) = V_{0}\cos(K x)$, $V_{0} = 2
a_{s1}/ a_{s0}$. For small values of $V_0$, the solution resembles
the unperturbed soliton with modulated parameters in the early
step of the propagation. This is true as long as the radiative
emission of matter wave is negligible. Using the collective
coordinate ansatz, we get that the soliton mass is preserved,
while the soliton center $\zeta$ obeys the effective particle
equation $\zeta_{tt} = -\partial V/\partial\zeta$ starting from
$\zeta(0)=\zeta_0$, $ \zeta_t (0)=4 \mu_0$, where the
effective potential  is
$$
V(\zeta) = -A_{nl}\cos(K\zeta),\ A_{nl} =
\frac{2\pi}{3}\frac{V_{0}\nu K}{\sinh(\frac{\pi K }{4\nu})}\left[
1 + \frac{K^2}{16\nu^2} \right].
$$
Here $A_{nl}$ is the  barrier for the soliton moving in the
nonlinear periodic potential. Note that in comparison with the
linear periodic potential $V(x)u(x,t)$ in Eq.~(\ref{gp2}), the
influence of the nonlinear periodic potential on the soliton is
enhanced. For the broad soliton case $K/\nu \gg 1$,  the
enhancement factor $\alpha = A_{nl}/A_{l}$ is $K^2/12, \alpha >
1$. For the narrow soliton case $K/\nu \ll 1 $, the enhancement
factor is $\alpha = 4\nu^2/3$. The soliton is moving as a
classical particle, and it can  be trapped  at $\zeta = 2\pi n/K,
n=0,1,2... $. In the trapped regime the soliton performs the
oscillatory motion with the small oscillation frequency $\Omega =
\sqrt{A_{nl}K}$. The critical velocity for depinning of the
soliton starting from the minimum of the potential is $v_{dp}=
\sqrt{2A_{nl}}$.

 When the
soliton width is much larger than the period of the nonlinear
potential, i.e. $K / \nu \gg 1$,
the radiation emission phenomenon
can be divided into two time steps. First, the soliton emits a small
but quick burst of radiation which is trapped in the form of
soliton shape modulation. Second, the soliton continues to radiate
slowly on long time scales $\sim V_0^{-2}$, as we shall see below.
The first step can be described in the "renormalized particle
limit" \cite{Scharf}. The dressed solution of Eq.~(\ref{gp2}) can
be searched in the form $u = u_s(x,t)(1 + \chi(x,t))$. For $v < K$,
where $v = 4\mu_{0}$ is the soliton velocity,
the solution for $\chi$ is
$$
\chi(x,t) = V_{0}\left( \frac{\cos(Kx)}{K^2 - v^2}
-i\frac{v\sin(Kx)}{K(K^2 -v^2 )} \right)|u^{(s)}|^{2}.
$$
At $v \sim K$ the approximation used
for the derivation is violated, and the radiation emission
should be studied more carefully.

{\it Emission of waves by  soliton  in a nonlinear periodic
potential}. The soliton propagating under action of periodic
nonlinear potential emits matter wave radiation. When the
modulations are weak we can use  the perturbation theory based on
the Inverse Scattering Transform to calculate the radiation
emission \cite{Karpman}. By conservation of the total mass $N$ and
energy $H$,
$$
 N = \int_{-\infty}^{\infty}|u|^{2}dx, \hspace*{0.1in}
  H = \int_{-\infty}^{\infty}
|u_{x}|^{2} - (1 + \frac{V(x)}{2}) |u|^{4}dx,
$$
the soliton parameters during the time interval $\Delta T$ are
modified according to $ \Delta\nu = -F(\nu, \mu,\Delta T)$,
$\Delta\mu = -G (\nu, \mu,\Delta T)$, where
\begin{eqnarray*}
&&F(\nu, \mu,\Delta T) = \frac{1}{4}\int n(\lambda,\Delta
T)d\lambda,\\
&&G(\nu, \mu, \Delta T) = \frac{1}{8}\int(\frac{\lambda^2}{\mu\nu}+
\frac{\nu}{\mu} - \frac{\mu}{\nu})n(\lambda, \Delta T)d\lambda.
\end{eqnarray*}
Here $n(\lambda,\Delta T)$ is the emitted mass density
during the time interval $\Delta T$, and $k=2\lambda$
(resp.  $\omega
= 4\lambda^{2}$) is the wavenumber (resp.
the frequency) of the emitted radiation.
The mass (number of atoms) emitted by the soliton
is $N_{rad} = \int n(\lambda)d\lambda$.
When $V_{0} \ll 1$ and the propagation times are of order
$V_{0}^{-2}$ we can calculate the emitted mass density and
the evolution of the soliton parameters. Different regimes are possible.

{\it Regime 1.} If the modulation wavenumber $K$ is smaller that
$\nu_{0}^2/\mu_{0}$, then the radiative emission is
negligible for times of order $V_{0}^{-2}$. The soliton parameters are almost
constant in this regime.

\begin{figure}
\begin{center}
\begin{tabular}{c}
\includegraphics[width=5.0cm]{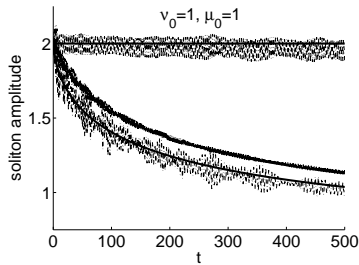}
\end{tabular}
\vspace*{-0.25in}
\end{center}
\caption{Soliton amplitude. The initial soliton parameters are
$\nu_0=\mu_0=1$. A periodic modulation $0.1 \cos(K x)$ is
applied to the nonlinear parameter. We compare the results from
full numerical simulations of the perturbed GP equation (thin
dashed lines) with the theoretical predictions of (\ref{sys1})
(thick solid lines). From top to bottom: $K=1.5$ (regime
1), $K=0.5$ (regime 3a), $K=1$ (regime 2a). The
strongest decay is achieved for the critical case
$K=1=\nu_0^2/\mu_0$, as predicted by the theory.
\label{fig1} }
\end{figure}

\begin{figure}
\begin{center}
\begin{tabular}{c}
\includegraphics[width=4.25cm]{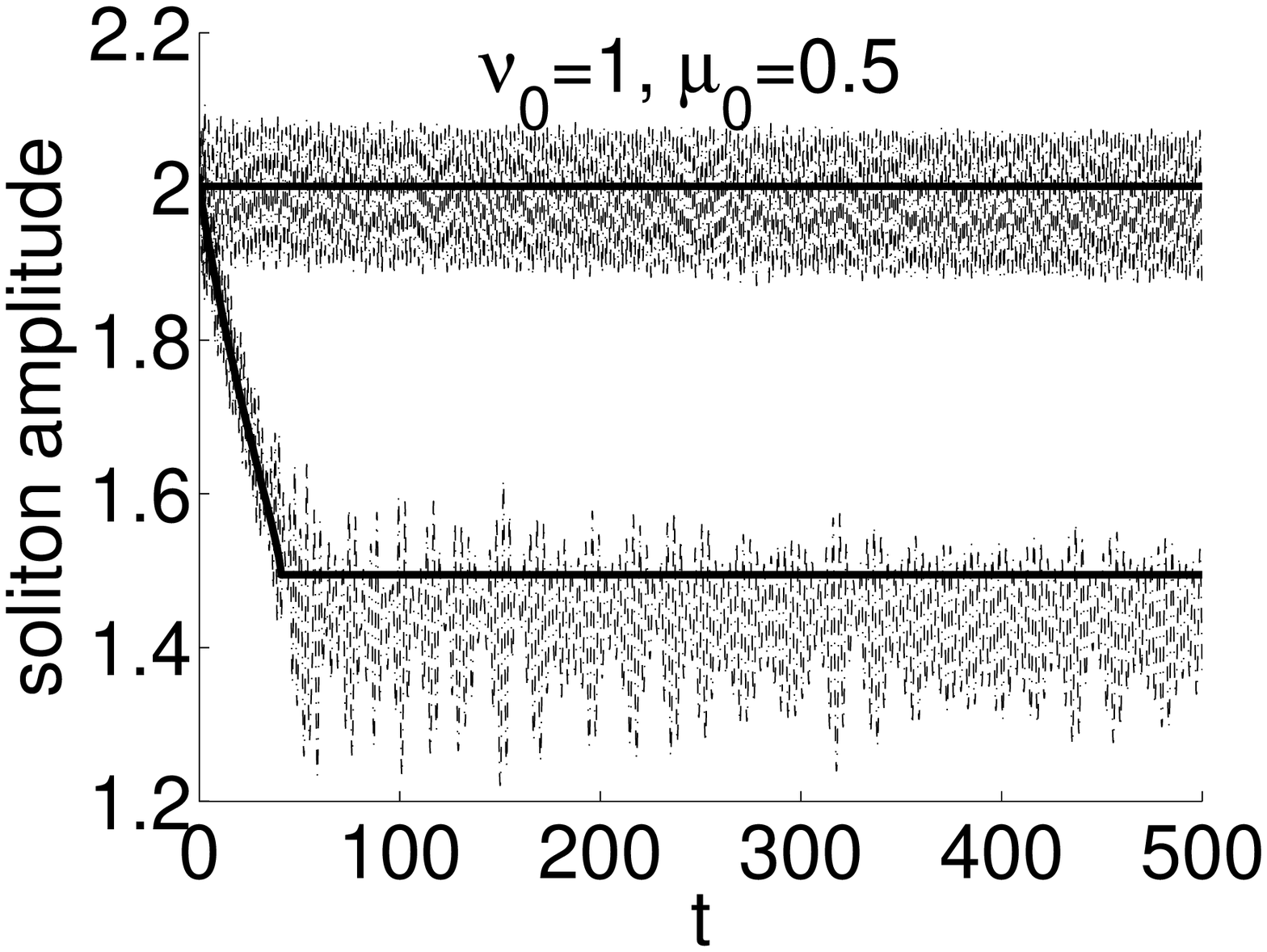}
\includegraphics[width=4.25cm]{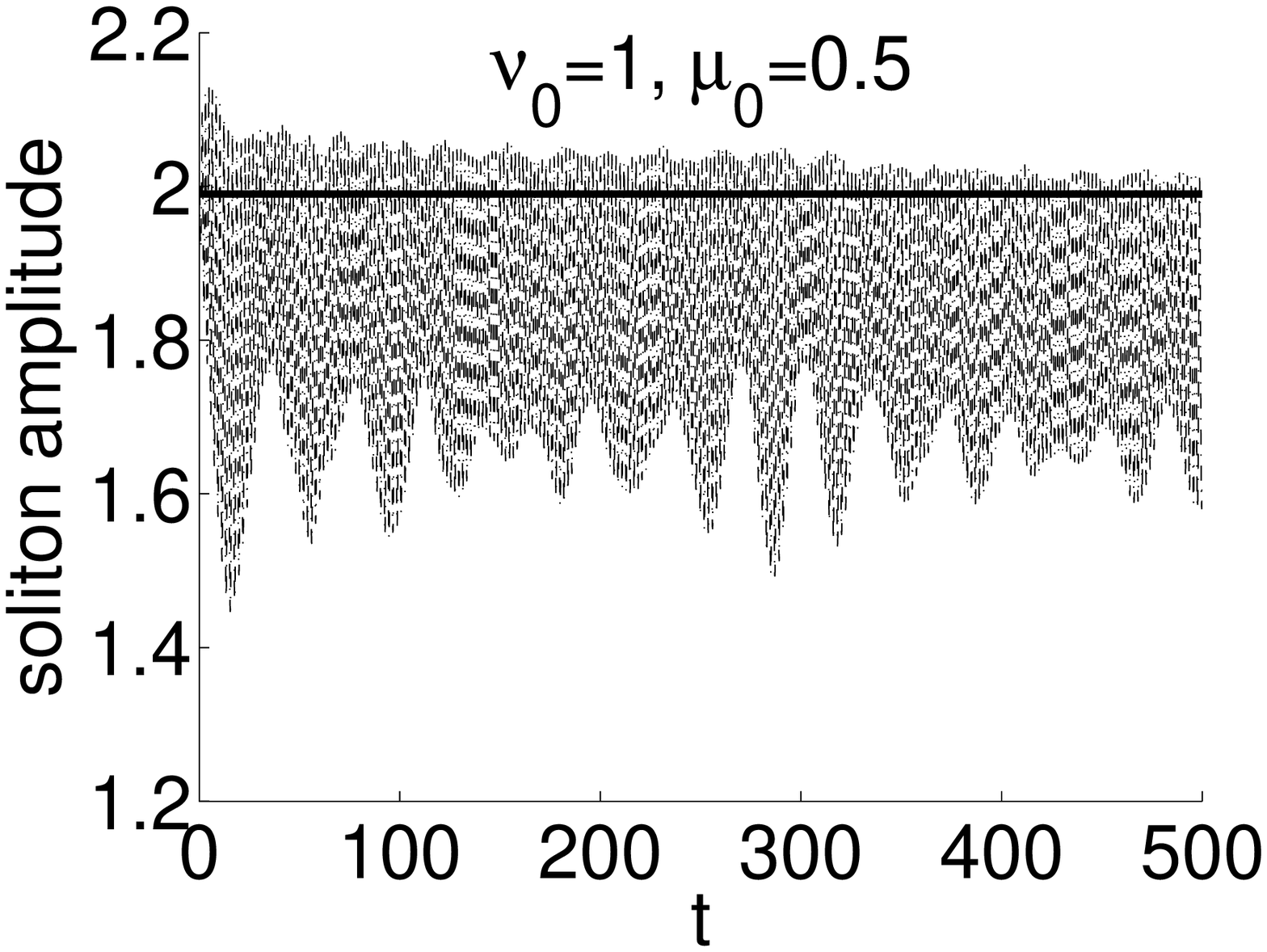}
\end{tabular}
\vspace*{-0.25in}
\end{center}
\caption{Soliton amplitude. The initial soliton parameters are
$\nu_0=1$, $\mu_0=0.5$. Left picture, from top to bottom:
$K=1.5$ (regime 1), $K=2.5$ (regime 2b). Right
picture:  $K=2$ (regime 3c). As predicted by the theory,
the critical case experiences oscillations but is stable.
\label{fig2} }
\end{figure}

\begin{figure}
\begin{center}
\begin{tabular}{c}
\includegraphics[width=4.25cm]{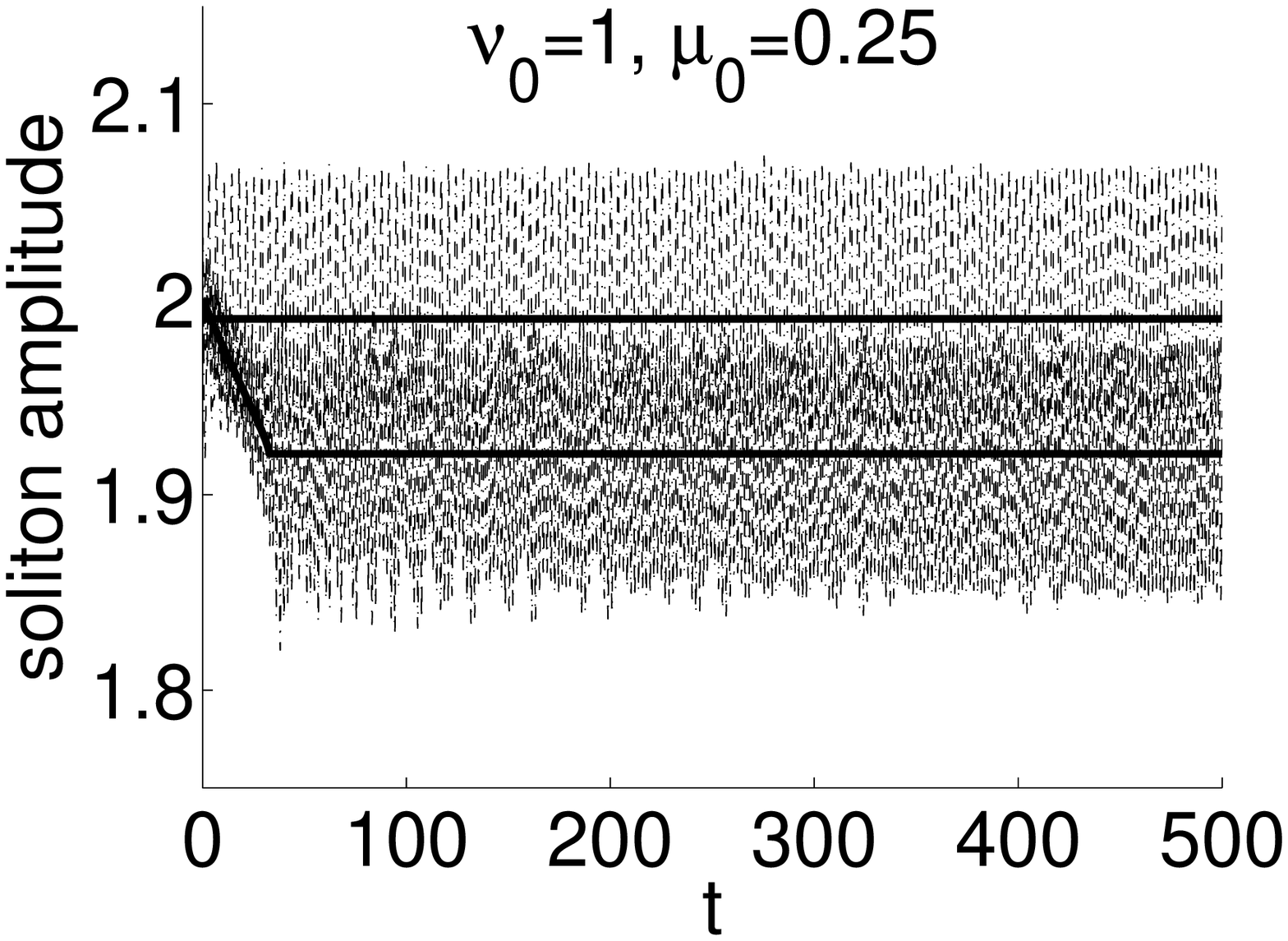}
\includegraphics[width=4.25cm]{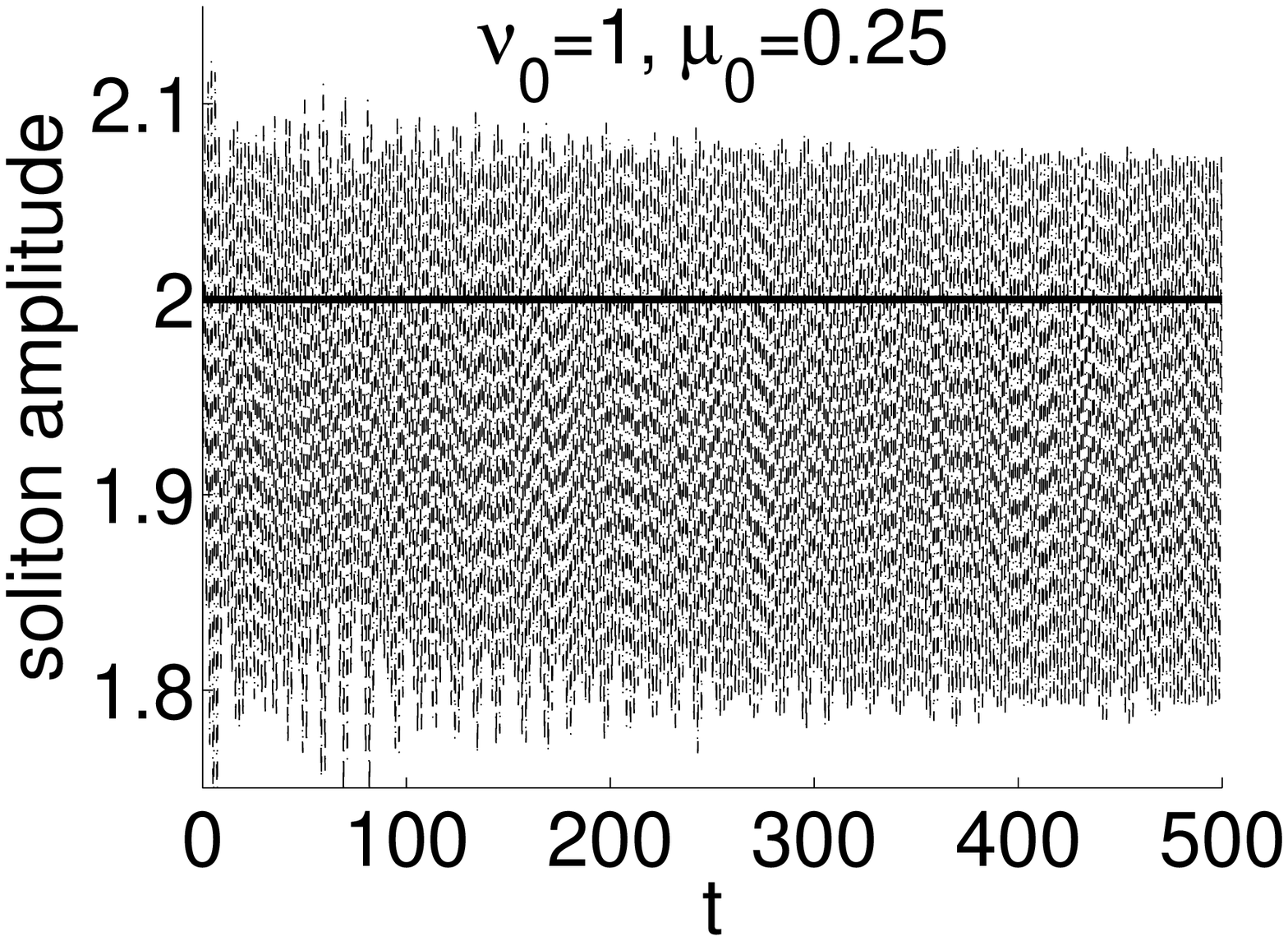}
\end{tabular}
\vspace*{-0.25in}
\end{center}
\caption{Soliton amplitude. The initial soliton parameters are
$\nu_0=1$, $\mu_0=0.25$. Left picture, from top to bottom:
$K=2$ (regime 1), $K=8$ (regime 2b). Right picture:
$K=4$ (regime 3b). As predicted by the theory, the critical
case is stable. \label{fig3} }
\end{figure}

{\it Regime 2.} If the modulation wavenumber $K$ is larger than
$\nu_{0}^2/\mu_{0}$, then the soliton emits a significant amount
of radiation. The soliton amplitude and velocity satisfy  the
system of ordinary differential equations
\begin{equation}
\label{sys1}
\nu_t = -F(\nu,\mu),\hspace*{0.1in}
 \mu_t = -G(\nu, \mu),
\end{equation}
starting from $\nu(0) = \nu_{0}, \mu(0) = \mu_{0}$. The functions
$F$ and $G$ are given by
\begin{eqnarray*}
F(\nu, \mu) = \frac{V_{0}^{2}}{16K\mu\lambda_+}
\left[\psi(\lambda_{+})^2 + \psi(\lambda_{-})^2 \right],
\\
G(\nu, \mu) = \frac{V_{0}^{2}}{16K\mu\nu\lambda_+}
\left[ \psi(\lambda_{+})^{2}(\frac{K}{2} + \lambda_{+}) \right.
\nonumber\\
\left. + \psi(\lambda_{-})^{2}(\frac{K}{2} + \lambda_{-}) \right],
\end{eqnarray*}
where $\lambda_{\pm} = \pm\sqrt{K\mu -\nu^2}$ and
\begin{eqnarray*}
\psi(\lambda) &=& \frac{\pi}{6\cosh( \pi (\lambda +
K/2)/(2\nu))}[ \nu^2 + (\lambda + \frac{K}{2})^2 ]\\
&&\times [\nu^2 + 2K \mu + (\lambda - \frac{K}{2})^{2}].
\end{eqnarray*}
The maximal radiative decay is obtained for $K$ close to $\nu^2/\mu$.
By integrating Eq.~(\ref{sys1}), we can put into evidence
that there are two subcases: {\it a).} The soliton mass $4\nu$ decays,
while the velocity increases or decays slowly, so that the soliton
stays in the regime $K\mu > \nu^2$.
Eq.~(\ref{sys1}) can be used  to describe
the long time behavior of the soliton, whose mass decays to zero
(Figure \ref{fig1}). {\it b).} The soliton mass $4\nu$ decays, but the
velocity $4\mu$ decays faster, so that the soliton parameters
reach the condition $K\mu = \nu^2$ at the critical time. This
state is a stable equilibrium, we recover the regime 1. This shows
that we can have a stable soliton even in the case $K >
\mu_{0}/\nu_{0}^2$, at the expense of the emission of a small
amount of radiation to allow the soliton to reach a stable state
(Figures \ref{fig2}-\ref{fig3}).

{\it Regime 3.} If we start from critical initial conditions such that
$\mu_{0}K = \nu_{0}^2$, then there are three sub-cases:
{\it a).} If $K <
4\mu_{0}$, then the soliton is attracted by the regime 2 and its
mass decays to 0. {\it b).} If $K > 4\mu_{0}$, then the soliton is
attracted by the regime 1 and its parameters remain constant.
{\it c).}
If $K = 4\mu_{0}$ (and thus $\nu_{0} = 2\mu_{0}$), the soliton
experiences strong oscillations but its mass does not decay
(Figure \ref{fig2}).

 {\it Emission of waves by  soliton  in a nonlinear random
potential}. Let us assume that the function $V$ is the
realization of a random zero-mean stationary process. The
correlation function is $B(x,l_{c}) = \left< V(x)V(0)\right>$, where $l_{c}$
is the correlation length. For times of order $B(0,l_c)^{-1}$
the soliton parameters satisfy the {\it
deterministic} system (\ref{sys1}) where the functions $F, G$ are
defined by
\begin{eqnarray*}
&&F(\nu, \mu) = \frac{1}{4\pi}\int_{-\infty}^{\infty} c^{2}(\nu , \mu, \lambda) d(k(\nu,\mu,\lambda))d\lambda,\\
&&G(\nu,\mu) = \int_{-\infty}^{\infty}\frac{\lambda^2 + \nu^2 -
\mu^2 }{8\pi\mu\nu } c^{2}(\nu, \mu, \lambda ) d(k(\nu, \mu,
\lambda )) d\lambda.
\end{eqnarray*}
Here the function $c$ is given by
\begin{eqnarray*}
c(\nu, \mu, \lambda)&=&
\frac{\pi}{96\mu^{9/2}}
(\nu^{2} + 17\mu^{2} -6\lambda\mu + \lambda^{2})\\
&&\times \frac{ ( (\lambda + \mu)^{2} +
\nu^{2} ) ((\lambda - \mu)^{2} +
\nu^{2} )}{\cosh(\pi(\nu^{2} + \lambda^2- \mu^{2} )/(4\mu\nu ))},
\end{eqnarray*}
and
the coefficients $d$ and $k$ by:
$$
k=\frac{(\lambda-\mu )^{2}+\nu^{2}}{\mu} , \hspace*{0.1in} d(k) =
2\int_0^{\infty}B(x)\cos(kx)dx.
$$
Note that $k(\lambda) \geq \nu^2/\mu$ for all $\lambda$.
Thus, the interaction between the soliton and the nonlinear random potential depends only the tail of the power spectral density $d(k)$ of $V$ for $k>\nu^2/\mu$.
There are two regimes of propagation:

{\it Regime 1.} If $\mu_0 \gg \nu_0$, then
the emitted radiation density is concentrated around the wavenumbers
$\pm 2\mu_0$. Besides the system (\ref{sys1}) can be simplified, and we obtain that the velocity of the soliton
is almost constant, while the mass decays as a power law:
\begin{equation}
\nu(t) \simeq \nu_0 \left(1 + \frac{t}{T_c} \right)^{-1/4},
\hspace*{0.1in} T_c=  \frac{3\mu_0 }{32 d(4 \mu_0) \nu_0^4}.
\end{equation}
In this regime the radiative decay prevents from transmitting
nonlinear wavepackets. The decay time is inversely
proportional to the correlation function of the disorder and the forth
power of the soliton amplitude. This means that this type of
disorder intensively destroys heavy solitons.

{\it Regime 2.} If $\mu_0 \ll \nu_0$, then
the soliton emits a very small amount of broadband radiation,
its mass is almost constant, while the velocity decays very slowly, typically
as a logarithm \cite{Garnier}. The amount of emitted radiation is proportional
to $d(\nu^2/\mu)$.
In this regime the soliton can be transmitted.

\begin{figure}
\begin{center}
\begin{tabular}{c}
\includegraphics[width=5.0cm]{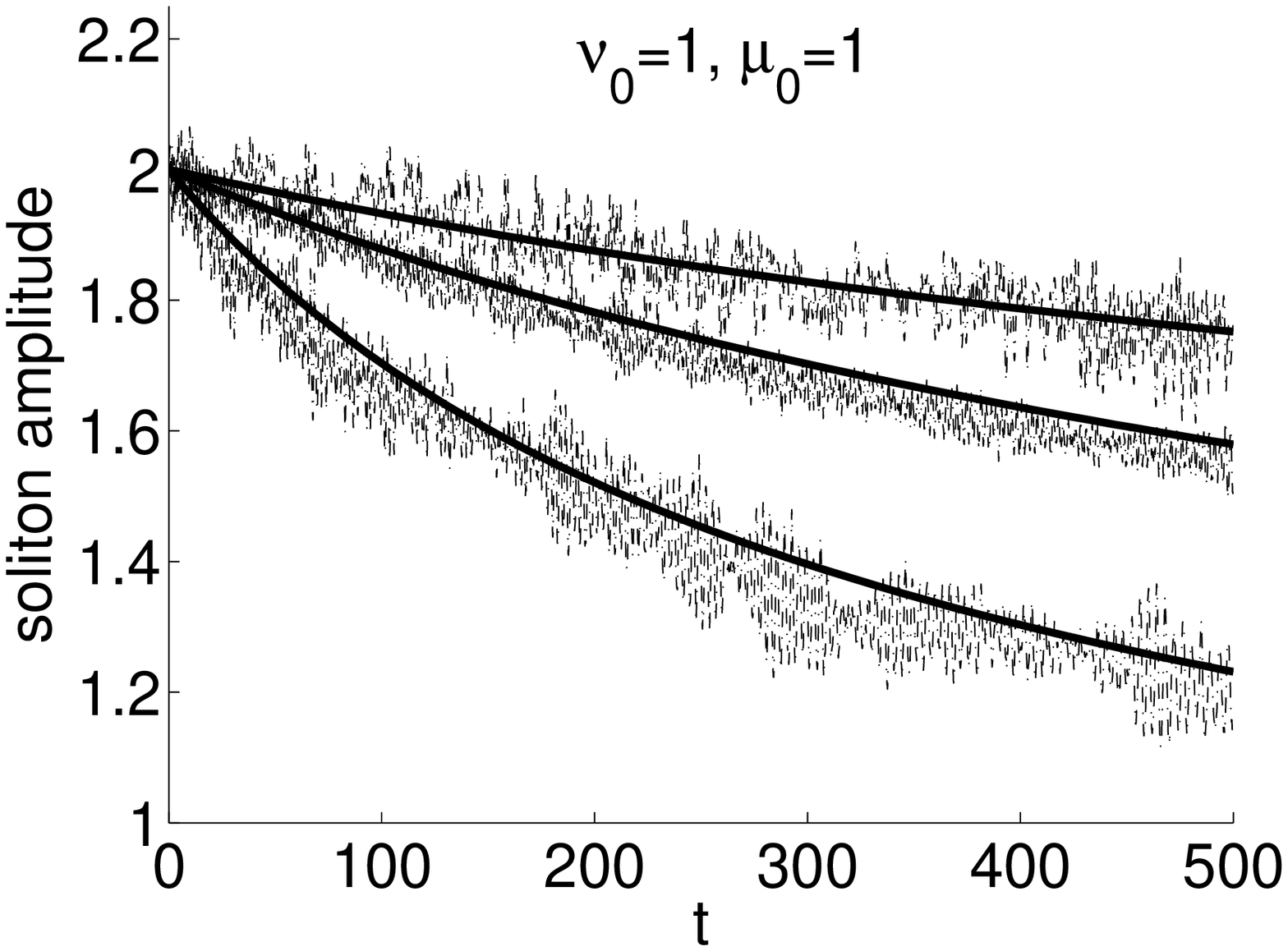}\\
\includegraphics[width=5.0cm]{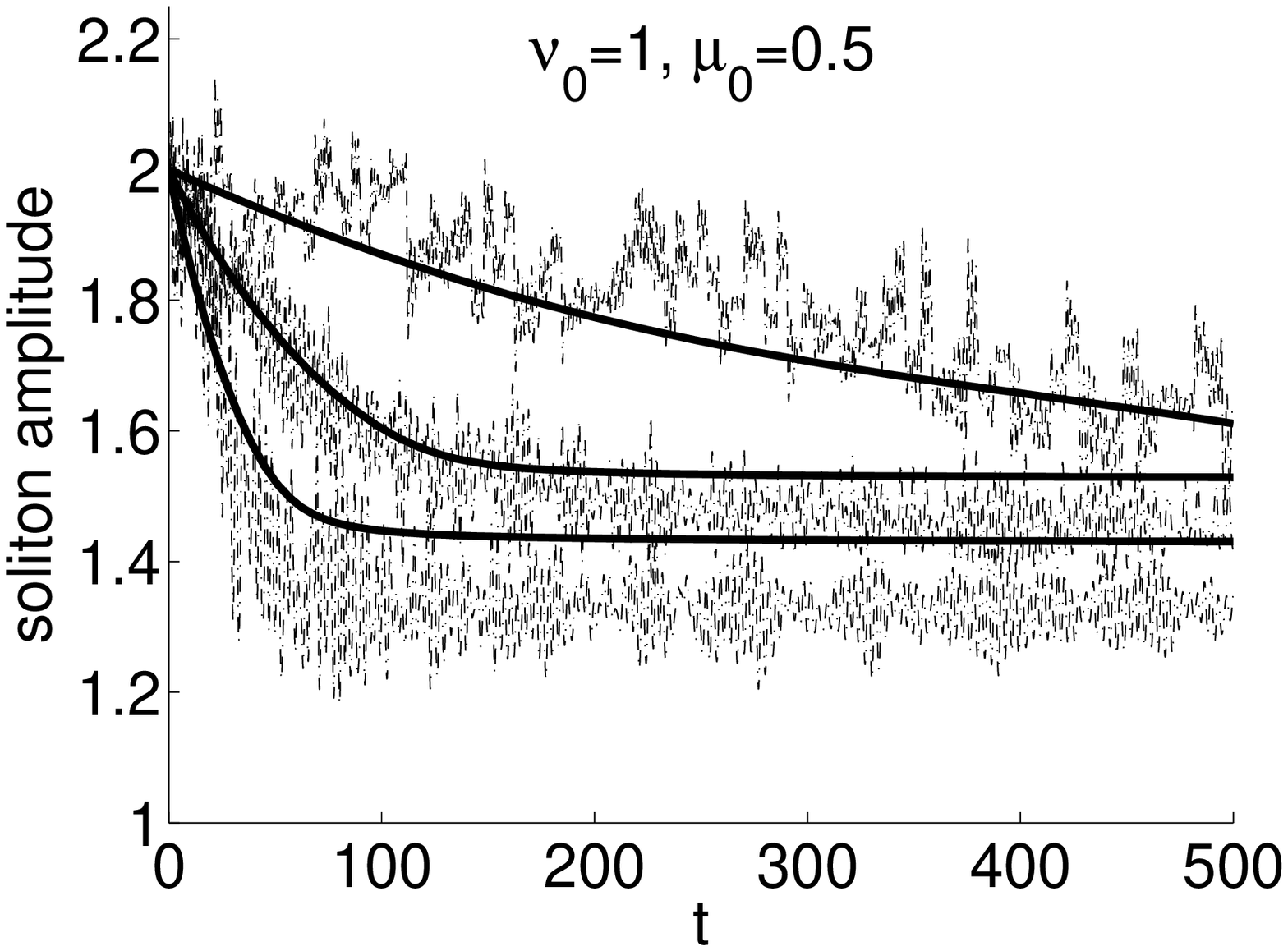}
\end{tabular}
\vspace*{-0.25in}
\end{center}
\caption{Soliton amplitude. A random modulation with amplitude $\sigma$
and correlation length $l_c$ is
applied to the nonlinear parameter.
The initial soliton parameters and potential amplitude are
$\nu_0=\mu_0=1$, $\sigma=0.1$ (top) and $\nu_0=1$, $\mu_0=0.5$
$\sigma=0.2$ (bottom).
 We compare the results from
full numerical simulations of the perturbed GP equation (thin dashed
lines) with the theoretical predictions of (\ref{sys1})
(thick solid lines). For each figure, we have from top to bottom: $l_c=8$,
$l_c=0.125$, $l_c=1$.
\label{fig4} }
\end{figure}

The  analysis of the system (\ref{sys1}) shows that these
two regimes are attractive,
in the sense that, after a transient regime, one observes the
regime 1 (resp. 2) if $\mu_0/\nu_0$ is above
(resp. below) a critical value.
We have checked these predictions by numerical simulations of
the randomly perturbed GP equation (\ref{gp2}).
We consider the case of a stepwise constant process
$V$. The constant step is equal to $l_c$, and the process
$V$ takes random independent values over each elementary interval
that are uniformly distributed in $[-\sigma,\sigma]$.
The power spectral density is $d(k) = 2 \sigma^2 [ 1 - \cos(k
l_c)]/[k^2 l_c]$. We compare in Figure \ref{fig4} the numerical
results and the theoretical predictions in a case close to the
regime 1 (top figure) and close to the regime 2 (bottom figure).
The power law radiative decay is noticeable in the top figure,
with a decay rate $T_c^{-1}$ that is maximal for $l_c\simeq
0.58$.
The full transmission regime (up to a
transient regime where the soliton emits radiation) can be
observed in the bottom figure.

 To estimate  the typical values of the parameters, let us consider
 the case of $^7$Li condensate with the FR at
$B = 720$G. The typical values of the scattering length used in
the experiments are: at $B=352$G is $a_s = -0.23$nm and
$a_{s}(300G) \approx -0.2$nm \cite{Straeker}. Thus by changing
periodically in space the magnetic field between these values we
can obtain the nonlinear periodic potential with the dimensionless
amplitude of modulations $V_{0} \approx 0.26$. In the trap with
$\omega_{\perp} = 2\pi\cdot 10^{3}$Hz and $n_{0} =
10^{6}$cm${}^{-1}$, we have $\xi \approx 2\mu$m, $c \approx 2$
mm/s, so $t_{0} \approx 1$ ms. A soliton with velocity $v \sim c$
travels through the region with modulated scattering length with
$L \sim 0.25$mm
 in the dimensionless time $t \sim 125$.  The
soliton width is of the order of $1.5\mu$m, and if the grating
period varies in the interval $(1,15)\mu $m, then both limits $K \nu
\gg 1$ and $K \nu \ll 1$ are covered. The optically induced FR
method gives a grating period $\sim 1\mu$m. The critical case $K=1=
\nu_{0}^2/\mu_{0}$ (Fig.~1) corresponds for these parameters to the
grating period $\approx 6\mu$m and the soliton velocity $\approx 8$mm/s.
The random modulations with the deviation
strength is $a_{s}/a_{s1} = 0.1$ can be achieved by the random
distribution of the current in wire along the atom
chip\cite{Gimperlein}. For the soliton velocity $v_{s} = 2c$, and
the initial number of atoms in soliton is $\sim 10^3$, we find
then the soliton decay time $T_{c} \sim 50$, which is equal to
$\sim 0.1$s in physical units.

{\it In conclusion} we have investigated the transmission of
matter wave bright solitons  through  periodic or random nonlinear
potential, generated by periodic or random spatial variations of
the atomic scattering length. The condition for the emission of
matter waves and radiative soliton decay are obtained. We show
that critical cases support the stable propagation of
bright solitons.

{\bf Acknowledgements.} The authors thank B.B. Baizakov and
M. Salerno  for interesting discussions. Partial financial support
of the work of F.Kh.A. from the Physics Department of Salerno
University is acknowledged.

\end{document}